\documentclass[a4paper,12pt]{article} 
\usepackage{amsmath}
\usepackage{amssymb,amsfonts,amsthm}
\usepackage{gastex}
\usepackage{mathtext}
\usepackage[T2A]{fontenc}
\usepackage[cp1251]{inputenc}
\usepackage[russian,english]{babel}
\usepackage{graphicx}
\usepackage{cite}               
\begin{document}

\newtheoremstyle{plain}{5pt}{5pt}{\normalfont\normalsize\itshape}{0pt}{\normalfont\normalsize\bfseries}{}{3pt}{{\thmname{#1}\thmnumber{\ #2}\thmnote{\rm #3}}}

\theoremstyle{plain}
\newtheorem{teo}{\indent Theorem}[section]
\newtheorem{cor}{\indent Corollary}[section]
\newtheorem{prp}{\indent Proposition}[section]
\newtheorem{conj}{\indent Conjecture}[section]

\renewcommand{\theteo}{\arabic{teo}}
\renewcommand{\thecor}{\arabic{cor}}
\renewcommand{\theprp}{\arabic{prp}}
\renewcommand{\theconj}{\arabic{conj}}
\renewcommand{\thetable}{\arabic{table}}
\renewcommand{\thefigure}{\arabic{figure}}
\renewcommand{\theequation}{\arabic{equation}}

\let\un=\underline
\let\ov=\overline
\let\ola=\overleftarrow
\renewcommand{\le}{\leqslant}
\renewcommand{\ge}{\geqslant}

\renewcommand{\exp}{{\sf exp}}
\def\cA{\mathcal{A}}
\def\cB{\mathcal{B}}
\def\cG{\mathcal{G}}
\def\cT{\mathcal{T}}
\def\cW{\mathcal{W}}
\def\Inc{{\sf Inc}}
\def\inc{{\sf Inc_{\prec}}}
\def\INC{{\sf \bar{I}nc}}
\def\CH{{\sf Ch}}
\def\OF{{\sf OF}}
\def\TM{{\sf TM}}
\def\RC{{\sf RC}}
\def\LC{{\sf LC}}
\def\SC{{\sf SC}}
\def\PM{{\sf PM}}
\def\MX{{\sf MX}}
\def\RT{{\sf RT}}
\def\Ind{{\sf Ind}}
\def\per{{\sf per}}
\def\lexp{{\sf lexp}}
\def\lev{{\sf lev}}
\def\For{{\sf Forbidden}}
\def\Back{{\sf Back}}
\def\Last{{\sf Last}}
\def\f{{\sf f}}
\def\lf{{\sf lf}}
\def\Def{{\sf def}}
\def\por{{\sf deg}}
\def\L{{\sf L}}
\def\M{{\sf M}}
\def\P{{\sf P}}
\def\C{{\sf C}}
\def\IR{{\sf Gr}}
\def\PD{{\sf Pd}}
\def\Pr{{\sf Pr}}
\def\Lex{{\sf Lex}}
\def\Tr{{\sf Tr}}
\def\bl{{\sf bl}}
\def\pre{{\sf PRE}}
\def\suf{{\sf SUF}}
\def\fact{{\sf FACT}}
\def\re{{\sf re}}
\def\Le{{\sf le}}
\def\e{{\sf e}}
\def\AD{{\sf AD}}
\def\Min{{\sf Min}}
\def\trim{{\sf Trim}}
\def\u{\mathbf{t}}
\def\v{\bar{\mathbf{t}}}
\def\h{{\sf h}}
\def\t{{\sf t}}

\def\S{{\char159}}
\newcommand{\II}{
\mbox{
\unitlength=1pt
\begin{picture}(-2.5,12)(2.5,5)
\gasset{Nw=1.5,Nh=1.5,Nfill=y,AHnb=0}
\node(o1)(0,16){}
\node(p1)(0,10){}
\node(q1)(0,4){}
\drawedge(o1,p1){}
\drawedge(p1,q1){}
\end{picture} }}
\newcommand{\XX}{
\mbox{
\unitlength=1pt
\begin{picture}(12,12)(7.5,5)
\gasset{Nw=1.5,Nh=1.5,Nfill=y,AHnb=0}
\node(o1)(5,16){}
\node(o2)(10,16){}
\node(o3)(15,16){}
\node(o4)(20,16){}
\node(p1)(5,10){}
\node(p2)(10,10){}
\node(p3)(15,10){}
\node(p4)(20,10){}
\node(q1)(5,4){}
\node(q2)(10,4){}
\node(q3)(15,4){}
\node(q4)(20,4){}
\drawedge(o1,p2){}
\drawedge(p2,q1){}
\drawedge(o2,p1){}
\drawedge(p1,q2){}
\drawedge(o3,p4){}
\drawedge(p4,q3){}
\drawedge(o4,p3){}
\drawedge(p3,q4){}
\end{picture} }}

\thispagestyle{empty}
\title{Combinatorial Characterization\\ of Formal Languages}
\author{Arseny M. Shur\thanks{This is the English translation of the ``autoreferat'' (extended abstract) of the dissertation presented for the doctoral (final) degree in physics and mathematics. The Russian original is available at {\tt wwwrus.imm.uran.ru/C16/Diss/Z/Автореф\_Шур.pdf}. The full text of dissertation (also in Russian) is available from the author, mailto {\tt Arseny.Shur@usu.ru}. This English version slightly extends the Russian one, because at the moment we do not plan to translate the full dissertation. We replaced the reference list of the autoreferat with the full bibliography given in the dissertation. On the other hand, we omitted or shortened some formal sections of the original version.}\\
Department of Algebra and Discrete Mathematics\\Ural State University}
\date{Ekaterinburg, Russia, 2010}
\maketitle

\noindent \textbf{National Classification of Scientific Areas}:\\[2pt]
01.01.06 --- Mathematical Logic, Algebra, and Number Theory\\[3mm]
\textbf{Consultant}: Prof. Lev N. Shevrin\\[3mm]
\textbf{Opponents (Readers)}:\\[2pt]
Prof. Farid M. Ablaev, Prof. Leonid A. Bokut', Prof. Vladimir I. Trofimov.\\[3mm]
\textbf{External Review by}: Moscow State University (Department of Mathematical Logic and Theory of Algorithms)\\[3mm]
\textbf{Date of Defence}: November 23, 2010\\[3mm]
\textbf{Jury}: No. D\,004.006.03 at the Institute of Mathematics and Mechanics, Ural Branch of Russian Academy of Sciences\\[1mm]
\textbf{Head of the Jury}: Prof. Aleksandr A. Makhnev

\newpage

\section{General matter on the dissertation}

\subsection*{Introduction to the topic}

Theory of formal languages plays an important role in contemporary mathematics. It is closely connected to such fundamental disciplines as algebra, logic, and combinatorics. Also, theory of formal languages cannot be separated from automata theory which studies acceptors and transducers of languages. Many algorithmic problems are formulated or can be easily reformulated as problems about formal languages. Formal languages have a number of applications in computer science (programming languages and compilers, software and hardware verification, data compression, cryptography, computer graphics, etc), and also in linguistics (natural languages processing, computer analysis of semantics, machine translation, dictionaries) and biology (analysis of DNA sequences, structure of proteins, populational dynamics, neural nets, membrane computation).

Formal languages are studied from different points of view. We point out five approaches; the researches on formal languages often contain elements of different approaches. Within the \textit{algebraic approach}, operations on languages, equations in words and languages, morphisms, congruences, and identities are studied. Also, there  are some specific ``algebraic'' languages, e.\,g., the language of minimal terms of an arbitrary fixed algebra. By means of the \textit{logical approach}, formal languages are just formula sets!! of different logics (usually, of the FO or SO logic with some restrictions and/or extensions). So, the main task in the logical approach is to capture the properties of languages with logical formalism. Another approach is to study the languages by means of generating systems (such as grammars) and accepting or transducing machines. Within the \textit{structural approach}, the properties of words are analyzed. Thus, a language is considered as the set of words defined by a common structural property. Finally, decidability and computational complexity of algorithmic problems about words and languages are studied within the \textit{algorithmic approach}. Note that combinatorial methods are widely used in all approaches. 

\medskip
Five mentioned approaches have a key common point. All of them use a quantitative characteristic of a formal language called \textit{combinatorial complexity}\footnote{Different approaches use different terminology. The terminology we adopt here is consistent and hardly can be misunderstood.}. The combinatorial complexity of a language $L$ is the most natural counting function associated with $L$. This function returns the number of words in $L$ of length $n$ and is denoted by $C_L(n)$.

\begin{itemize}
\item 
In the study of algebras, it is often useful to estimate the \textit{growth} of an algebra, i.\,e., the combinatorial complexity of the language of minimal terms. Growth problems were studied for groups, semigroups, rings, modules, and some other types of algebras. A (far from complete) list of papers on this topic includes \cite{Bab,Bra,C-SG,Gov,Grg1,Grg,GrgH,Mi,Tro2,Uf1,Uf2,Wo}. The most remarkable result on the growth of groups is Gromov's theorem \cite{Gro} stating that a finitely generated group has polynomial growth if and only if it is nilpotent-by-finite. Concerning the growth of noncommutative algebras, we should mention the book by Krause and Lenagan \cite{KL}.
\item An important characteristic of a logical formula is the number of non-isomorphic finite models of given fixed type and given size. If the models are words, then one gets the combinatorial complexity of the language defined by the formula. This characteristic is often calculated for other types of models, e.\,g., for graphs, see the book \cite{GKLM}. We give just two examples. Fagin \cite{Fag} established that any set of graphs defined by a FO formula has either density 1 or density 0 in the set of all graphs. This paper started an extensive study of 0--1 laws on graphs. Second, we mention the investigations of the growth of hereditary (closed under generated subgraphs) classes of graphs. Such classes are direct analogues of factorial languages, which are quite common objects in the studies of combinatorial complexity. E.\,g., it is known \cite{BBW1,BBW2,SZ} that only six types of growth (constant, polynomial, exponential, and three factorial ones) are possible for hereditary classes of graphs\footnote{Logical approach to languages also generated another important quantitative characteristic. \textit{Descriptive complexity} of a language equals the size of the minimal model of a given type, generating the language. This characteristic is inspired by the notion of Kolmogorov complexity, see the book \cite{LV}.}. 
\item Grammars are also closely connected to combinatorial complexity. Chomsky and Schutzenberger established \cite{CS} that if a language is regular (i.\,e. is generated by a right-linear grammar), then its combinatorial complexity satisfies some linear homogeneous recurrence relation with constant coefficients and thus has rational generating function. Further, they proved that generating function of the combinatorial complexity of any unambiguous context-free language is algebraic. The latter result was later completed by Flajolet \cite{Fla} who showed that such a generating function for an ambiguous context-free language can be transcedental\footnote{A remarkable result in the converse direction was proved in 2010 within the \textit{logical} approach: any function $f:\mathbb{N}_0\to\mathbb{N}_0$ satisfying a linear homogeneous recurrence relation with constant coefficients equals the difference of combinatorial complexities of two regular languages \cite{KM}.}. 
\item Within the structural approach, the \textit{subword complexity} functions are studied for infinite words. Subword complexity is just the combinatorial complexity of the set of finite factors of an infinite word\footnote{For infinite words, a \textit{topological} approach is also quite useful. This approach includes the study of the function which is similar to subword complexity and is called an \textit{entropy} of an infinite word.}. The first results on subword complexity (and in fact, on combinatorial complexity at all) were obtained by Morse and Hedlund in 1938--1940 \cite{MH,MH2}. A systematic study of subword complexity was initiated by Ehrenfeucht and G. Rozenberg, see \cite{ERL1,ERL2,ERL3,ER1,ER2,ER3,ER4,ER5,ER6}. We point out a nice classification of morphisms w.\,r.\,t. subword complexity of their fixed points, given by Pansiot \cite{Pan2}. In addition, there are several other counting functions on infinite words, such as palindromic, arithmetic, pattern, maximal pattern, and permutational complexities, see \cite{ABCD,ACF,AFF,Frid,KZ1,KZ2,Mak,RS,Wi}.
\item Except for the simple fact that the cost of brute force search algorithms depends on the size of the searched language, the connection of algorithmic approach to combinatorial complexity is not so obvious.  A nontrivial example of such a connection is given in the dissertation.
\end{itemize}

During the last decades, a lot of papers on combinatorial complexity was published. A deep study of subword complexity of infinite words (in addition to the above references, see \cite{All,AS,Brl,Cas2,Cas3,Cas4,CH,Dev,Frid1,Frid2,FA,Gri}) resulted in satisfactory answers to the most natural questions. Besides the infinite words, most papers about combinatorial complexity concern just a single language each, see \cite{ACR,BCJ,Car98,Cas1,Cur04,Edl,Grim,JPB,RS1,KS,Ko2,Kol,Kol07,Lep,OR,RG}. The other results, see \cite{BG,BB,C-SW1,C-SW2,C-S,CS,AIV,GKRS,Tro}, look rather scattered. There is a certain need in some unified theory that 
\begin{itemize}
\item explains connections between the structure and the growth properties of a language,
\item provides algorithms and formulas to find or approximate the parameters of growth of a language,
\item predicts the impact of the variations in the properties defining a language on its combinatorial complexity.
\end{itemize}
We are going to make some steps towards the construction of such a theory. In order to do this, we developed the following program.

\subsection*{Research program}

\begin{figure}[!htb]
\centerline{\includegraphics{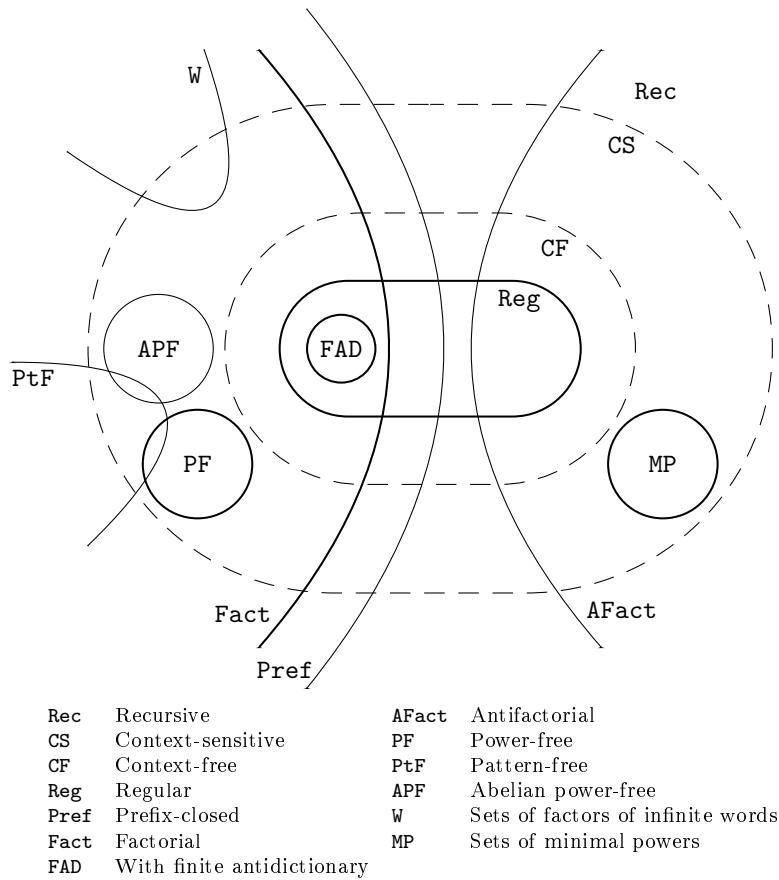}}
\caption{\small\sl Classes of languages considered in the dissertation. Main [resp., secondary] objects of study are drawn by thick [resp., thin] lines. The two middle classes of Chomsky hierarchy are drawn by dashes to indicate that we do not study them in general.}\label{obj}
\vspace*{-1.5mm}
\end{figure}

In what follows, ``complexity'' of a language \textit{always means combinatorial complexity}. To study the complexity of a language $L$ we should have an algorithm deciding whether $w\in L$ for any given word $w$. The existence of such an algorithm means exactly that $L$ is \textit{recursive}. So, all further considerations are within the class $\tt Rec$ of recursive languages. When we speak about a class of languages, we mean the intersection of this class with $\tt Rec$. Classes of languages considered in the dissertation are presented in Fig.~\ref{obj}. The main objects of study are marked in this figure by thick lines.

Studying a language, we are interesting in the asymptotic behaviour of complexity rather than the precise values of it. In particular, we consider finite languages as a degenerate case. Thus, ``degenerate'' intersections of the classes of languages are not represented in Fig.~\ref{obj}. The main parameter of the asymptotic behaviour of complexity is the \textit{growth rate} of a language $\IR(L)=\varlimsup\limits_{n\to\infty}(\C_L(n))^{1/n}$.
To compare functions, we use the standard $O$, $\Omega$, and $\Theta$ notation.

\begin{itemize}
\item \textit{Regular languages} constitute one of the most important classes of languages and have a number of equivalent definitions (defined by regular expressions, recognized by finite monoids, expressed in monadic SO logic, generated by right-linear grammars, recognized by finite automata, and so on). The main theorem on the complexity of regular languages can be obtained putting together several results from the book by A. Salomaa and Soittola \cite{SS}. Slightly simplifying, we can state this theorem as follows. For each regular language $L$ there is a number $r\in\mathbb{N}$, and for each $j=0,\ldots,r{-}1$ there exist a real polynomial $p_j(n)$ and algebraic real numbers $\alpha_j$, $\gamma_j$ such that $\alpha_j=\gamma_j=0$ or $0\le\gamma_j<\alpha_j$, and
\begin{equation}
\C_L(n)=p_{n'}(n)\alpha_{n'}^n+O(\gamma_{n'}^n),\text{\ where\ }n'=n\bmod r. \label{saso}
\end{equation}
The only but significant disadvantage of the above description is the lack of connection between the description and the properties of the language $L$ (or the parameters of the construction which defines $L$). As a result, no efficient algorithms to calculate the asymptotic parameters of $\C_L(n)$ were known except for the folklore algorithm to calculate $\IR(L)$ (i.\,e., the maximum of the numbers $\alpha_j$). This algorithm is polynomial but not enough efficient for practical calculation. Thus, considering deterministic finite automata (dfa's) as the most convenient and natural way to represent regular languages, it is natural to state the following problems.

\begin{itemize}
\item [{\tt Reg1}:] for dfa's, describe the properties that are responsible for the parameters of asymptotic behaviour of the complexity of corresponding regular languages;
\item [{\tt Reg2}:] describe possible oscillations of complexity for regular languages;
\item [{\tt Reg3}:] find an efficient algorithm to calculate, up to the $\Theta$-class, the complexity of a language from a dfa recognizing this language. 
\end{itemize}
\item \textit{Factorial languages} are the languages closed under taking factors of words. The class of factorial languages is wide. In particular, it contains the languages of minimal terms of algebras, the languages of factors of infinite words, and the languages defined by avoidance properties of words. The \textit{antidictionary} of a factorial language $L$ consists of all words that are minimal w.\,r.\,t. the containment order among the words from the complement of $L$\footnote{The complement of $L$ is an ideal of the free monoid over the alphabet of $L$, and the antidictionary is the minimal generating set of this ideal.}. Clearly, $L$ is determined by its antidictionary.

In \cite{ER4,BB}, factorial languages of bounded complexity are studied. No general results on the complexity of factorial languages are known. To study the complexity of factorial languages, it is convenient to use \textit{method of regular approximations}, described in the second part of this paper. This method uses the regular languages with the same local structure of words as in the target language. The following problems arise.
\begin{itemize}
\item [{\tt Fact1}:] study the convergence of the method of regular approximations and the restrictions on the use of this method;
\item [{\tt Fact2}:] find a nontrivial example of factorial language such that the exact growth rates can be found for all regular approximations of this language;
\item [{\tt Fact3}:] find language transformations preserving growth rates of factorial languages.
\end{itemize}
\item \textit{Languages with finite antidictionary} (FAD-languages) are contained in the intersection of the two previous classes. These are exactly the languages that serve as regular approximations of factorial languages. In most cases FAD-languages are given by their antidictionaries. We mention the Goulden--Jackson cluster method \cite{GJ,NZ} to build the generating function for the complexity of any FAD-language from its antidictionary. But this method is too time-consuming to process big antidictionaries, and hence, to obtain sharp bounds for the complexity of factorial languages through their regular approximations. The following problems arise naturally.
\begin{itemize}
\item [{\tt FAD1}:] find all, up to the $\Theta$-class, possible complexities of FAD-languages;
\item [{\tt FAD2}:] find, which transformations of antidictionaries preserve the asymptotic parameters of complexity of FAD-languages;
\item [{\tt FAD3}:] characterize the dfa's recognizing FAD-languages.
\end{itemize}
\item \textit{Power-free languages} constitute a well-known class of factorial languages. They were extensively studied since the seminal papers by Thue \cite{Th06,Th12}. Let $w$ be a word of length $n$, and let $\beta>1$. The $\beta$-\textit{power} of $w$ is the word
$$
w^{\beta}=\underbrace{w\ldots w}_{\lfloor\beta\rfloor\,\text{times}}w'\text{\ of length\ }\lceil\beta n\rceil,\text{ where }w'\text{\ is a prefix of\ }w.
$$
A word is $\beta$-\textit{free} $[\beta^+$-\textit{free}$]$ if it contains no $\beta$-powers $[$resp., no $\beta'$-powers satisfying $\beta'>\beta]$. The language $\L(k,\beta)$ $[\L(k,\beta^+)]$ consists of all $\beta$-free $[$resp.,  $\beta^+$-free$]$ words over the $k$-letter alphabet\footnote{It is convenient to consider  $\beta^+$ as a <<number>> such that the inequalities $x\le\beta$ and $x<\beta^+$ are equivalent. Once the set of powers is extended in this way, we use only the notation $\L(k,\beta)$.}. For a fixed alphabet, the size of a power-free language grows as $\beta$ increases. Hence, there exists \textit{repetition threshold} $\RT(k)$ separating finite and infinite $k$-ary power-free languages. The values of $\RT(k)$ were conjectured by Dejean in 1972 \cite{Dej}. Namely, $\RT(3)=\frac{7}{4}$, $\RT(4)=\frac{7}{5}$, and $\RT(k)=\frac{k}{k{-}1}$ otherwise. The proof of Dejean's conjecture was finished in 2009, see \cite{Rao,Mou,CR1,CR2,Pan1,Car2,MNC}. The known results on complexity of power-free languages are related to a few particular languages, see the survey \cite{Ber}. More than ten papers were devoted to the growth rate of the language $\L(3,2)$: the best upper bound was obtained by Ochem and Reix \cite{OR}, and the best lower bound was given by Kolpakov \cite{Kol07}. The most interesting feature found so far is the ``polynomial plateau'' of complexity, discovered by Karhum\"aki and Shallit \cite{KS} in the binary case: all power-free languages between $\L(2,2^+)$ and $\L(2,7/3)$ have polynomial complexity (and, moreover, quite close orders of polynomial growth). The complexity of $\L(2,2^+)$ was estimated with increasing precision in \cite{RS1,Lep,Ko2,Cas1}; the final result was obtained in \cite{JPB}. The following problems on power-free languages should be considered.
\begin{itemize}
\item [{\tt PF1}:] find a property (of powers) that can explain the existence of the polynomial plateau;
\item [{\tt PF2}:] prove a connection between low combinatorial and low computational complexity, solving the \textit{context equivalence problem}\footnote{This problem, which is a version of the word problem, will be introduced in the second part of this paper.} for the language $\L(2,2^+)$ from the polynomial plateau;
\item [{\tt PF3}:] build universal algorithms to estimate the growth rates of power-free languages both from above and from below;
\item [{\tt PF4}:] describe the growth rate of the languages $\L(k,\beta)$ as a function of $k$ and $\beta$;
\item [{\tt PF5}:] describe structural properties of the minimal infinite power-free languages over different alphabets (\textit{threshold} languages).
\end{itemize}
\item \textbf{Languages of minimal powers} are exactly the antidictionaries of power-free languages. Any such language is \textit{antifactorial}, i.\,e. constitutes an antichain w.\,r.\,t. the containment order. Antifactorial languages are closely connected to the factorial ones, but their complexity behaves irregularly and is completely unexplored. Here we state only one problem; it significantly generalizes Problem~1.12 of \cite{AS}.
\begin{itemize}
\item [{\tt MP}:] for any language of minimal powers, describe the set of zeroes of its complexity. 
\end{itemize}
\end{itemize}

\subsection*{Aim of dissertation}
The dissertation is aimed at the development of new approaches to study combinatorial complexity. We apply these approaches to different classes of formal languages
\begin{itemize}
\item to discover and estimate the impact of the properties of languages (and of the structures associated with languages) on complexity;
\item to provide algorithms and formulas estimating the complexity for wide classes of languages; 
\item to discover connections between combinatorial and computational complexity.
\end{itemize}
Particular goals of the dissertation are solutions to the fifteen problems mentioned in the research program.

\subsection*{Methods}
The methods used in the proofs of the obtained results can be grouped as follows.

\begin{itemize}
\item Methods of \textit{combinatorics of words}, based on the properties of periodic words, the properties of Thue-Morse words and Thue-Morse morphism, construction and analysis of morphisms, encodings, circular and two-dimensional words.
\item Methods of \textit{automata theory}, in particular, original construction of \textit{web-like} and \textit{generalized web-like automata}. This construction allows us to prove several theorems which have nothing in common at first glance.
\item Methods of \textit{matrix theory}, based on the Perron-Frobenius theorem and related properties of nonnegative matrices. We also use the Jordan normal form, the Hamilton-Cayley theorem, and the calculation of determinants of variable size. 
\item Methods of \textit{graph theory}, including equitable partitions, analysis of strongly connected components, and some spectral properties of graphs.
\end{itemize}

We also make use of some classical combinatorial algorithms such as Tarjan's algorithm for finding strong components of a digraph and Aho--Corasick's algorithm for pattern matching. Finally, we use computer to calculate numerical bounds of complexity and also to search examples and make routine computations in some proofs.

\subsection*{Size and structure of dissertation. Publications}
The dissertation (287 pages) consists of introduction (Sect. $1^{\circ}\!{-}3^{\circ}$), four chapters (\S\S\,1--20), bibliography, and index. The results constituting the dissertation are published in the papers \cite{SSh,Sh_r98,Sh_r00,Sh_r05,Sh_r09iu,Sh_r09iv,Sh_r10ti,Sh_r10dr,Sh96ac,Sh96sf,Sh06dlt,Sh06jm,Sh07,Sh08csr,Sh08ita,ShGo08,Sh09ita,Sh09dam,Sh09dlt,
ShGo10,Sh10tcs,Sh10csr,Sh10dlt,Sh10ejc,SaSh}. In addition, the manuscripts \cite{Sh11sf,Sh11_1,Sh11_2,Sh11_3} are submitted or will be submitted soon. The papers \cite{Sh08ita,ShGo10,Sh10tcs,Sh10csr} are the extended versions of \cite{Sh06jm,ShGo08,Sh_r09iv,Sh_r10dr}, respectively.

\subsection*{Acknowledgements}
With a great pleasure I express my gratitude to Lev Shevrin for his permanent attention to this work and lots of remarks on the draft version of the dissertation. I should mention my first supervisor Evgeny Sukhanov, who taught me the basics of formal language theory and, what is even more important, the basics of research work. I am grateful to Mikhail Volkov, Andrei Bulatov, and Juhani Karhum\"aki for useful and stimulating discussions. Special thanks to my students Irina Gorbunova, Alexey Samsonov, and Anastasia Tkach, who wrote computer impementations of several algorithms proposed in the dissertation, thus making possible extensive numerical studies.

\section{Results}

We solved, completely or partially, all fifteen problems mentioned in the research program. The results are highly connected with each other: more than twenty statements are used outside the sections when they were proved. These connections witness the possibility to build a unified theory of combinatorial complexity. Now we start to describe the results.

\subsection*{Chapter 1 (\S\S\,1--4). Regular languages}

According to the formula (\ref{saso}) given above, the asymptotics of complexity is fully described by the set of $r$ \textit{asymptotic functions} $p_j(n)\alpha_j^n$ (or, up to the $\Theta$-equivalence, $n^{m_j}\alpha_j^n$). The parameter $\alpha_j$ of the fastest growing asymptotic function equals $\IR(L)$, while the parameter $m_j=\PD(L)$ is the so-called \textit{polynomial index} of $L$\footnote{Polynomial index can be defined by the inequality $0<\limsup\limits_{n\to\infty}\dfrac{C_L(n)}{n^{\PD(L)}\IR(L)^n}<\infty$. Besides the class of regular languages, we consider polynomial indices only for the languages of polynomial complexity.}.

We consider finite automata as labeled digraphs. A dfa is \textit{consistent} if for any of its states (vertices) there is an accepting path containing this state. The proofs of most statements in Chapter 1 result from the structure and mutual location of \textit{strong components} (i.\,e., maximal strongly connected subgraphs) of consistent dfa's. First we give the polynomiality criterion.

\begin{teo}[\cite{Sh_r05}] \label{reg1} 
Suppose that a language $L$ is recognized by a consistent dfa $\cA$. Then\\
$(1)$\,if $\cA$ is acyclic, then $L$ is finite;\\
$(2)$\,if $\cA$ contains two cycles sharing a common vertex, then  $L$ has exponential complexity;\\
$(3)$\,if $\cA$ contains cycles but any two of them are disjoint, then $L$ has polynomial complexity and $\PD(L)=m{-}1$, where $m$ is the maximum number of cycles intersected by a single walk in $\cA$. 
\end{teo} 

\begin{cor}[\cite{Sh_r05}] \label{creg1}
If a regular language $L$ over $k$ letters is recognized by a consistent dfa with $N$ vertices, then it is decidable in $O(Nk)$ time whether the complexity of $L$ is polynomial or exponential. In the first case, the polynomial index of $L$ can be found in time $O(Nk)$ also.
\end{cor}

Next, consider the problem of finding the growth rate $\IR(L)$. Recall that the \textit{index} $\Ind(G)$ of a graph $G$ is the Frobenius root\footnote{The Frobenius root, i.\,e., the maximal in absolute value eigenvalue of a non-negative matrix, is one of the most important spectral characteristic of a graph.} of the adjacency matrix of $G$. A folklore result says that $\IR(L)=\Ind(\cA)$ for any consistent dfa $\cA$ recognizing $L$. In general, the Frobenius root of an adjacency matrix cannot be found exactly but can be approximated with the absolute error $\delta$ for any $\delta>0$. A straightforward computation uses characteristic polynomial of the matrix and requires $\Omega(N^4)$ operations and $\Omega(N^3)$ additional space. The following theorem radically improve this situation.

\begin{teo}[\cite{Sh08csr,Sh10tcs}] \label{reg3} 
Suppose that a language $L$ over $k$ letters is recognized by a consistent dfa $\cA$ with $N$ vertices. There is an algorithm which, given $\cal A$ and a number $\delta$, $0<\delta<1$, calculates $\IR(L)$ with the absolute error at most $\delta$ in time $\Theta(\log(1/\delta){\cdot}Nk)$ using $\Theta(\log(1/\delta){\cdot}N)$ additional space.
\end{teo}

The mentioned algorithm (\textbf{Algorithm R}, \cite{Sh10tcs}) plays an important role in the dissertation. Note that it can be used to calculate the index of any graph, as the following theorem shows.

\begin{teo}[\cite{Sh10tcs}] \label{reggra}
Let $G$ be a digraph with $n$ vertices and $m$ edges. There is an algorithm which, given $G$ and a number $\delta$, $0<\delta<1$, calculates $\Ind(G)$ with the absolute error at most $\delta$ in time $\Theta(\log(1/\delta){\cdot}m)$ using $\Theta(\log(1/\delta){\cdot}n)$ additional space.
\end{teo}

Next we estimate the number of asymptotic functions, using the following technical notion. A strong component $C$ of a consistent dfa $\cA$ is \textit{important} if there is infinitely many numbers $n_i$ such that (a) there is an accepting walk of length $n_i$ intersecting $C$ and (b) there is no accepting walk of length $n_i$ intersecting a strong component with the index greater than $\Ind(C)$.
Recall that \textit{imprimitivity number} of a digraph is the greatest common divisor of the lengths of all its cycles. To obtain the following theorem we give a direct proof of the formula (1) by means of matrix theory.

\begin{teo}[\cite{Sh_r10ti}] \label{reg4}
Suppose that a language $L$ is recognized by a consistent dfa $\cA$, $r$ is the least common multiple of the imprimitivity numbers of all important strong components of $\cA$. Then the complexity of $L$ can be described by $r$ asymptotic functions. 
\end{teo}

The following theorem describes the polynomial index of a regular language in the general case.

\begin{teo}[\cite{Sh08csr,Sh_r10ti}] \label{reg5}
Suppose that a language $L$ is recognized by a consistent dfa $\cA$, $m$ is the maximum number of strong components of index $\IR(L)$ intersected by a single walk in $\cA$. Then $\PD(L)={m{-}1}$.
\end{teo}

To calculate polynomial index, we need to prove or disprove the equality of indices of two digraphs in the case when these indices are equal up to the approximation error. 

\begin{prp}[\cite{Sh_r10ti}] \label{reg6} 
Let $\mathbb{A}_{k,N}$ be the set of all consistent dfa's having at most $N$ vertices and acting over the $k$-letter alphabet. If $\cA,\cB\in\mathbb{A}_{k,N}$, then the equality of the numbers $\Ind(\cA)$ and $\Ind(\cB)$ can be verified in time $O(N^4 +\log(1/\delta(N)){\cdot}N^2)$, where $\delta(N)$ is the minimum nonzero difference of indices of two dfa's from $\mathbb{A}_{k,N}$.
\end{prp}

The proof of Theorem~\ref{reg4} provides a way to get the parameter $r$ and allows one to reduce the calculation of the numbers $\alpha_j$, $m_j$ for $j=0,\ldots,r{-}1$ to the calculation of the growth rates and polynomial indices of some subgraphs of the dfa $\cA$. Thus, Problems {\tt Reg1} and {\tt Reg3} are completely solved.

\smallskip
The function $f$ is called \textit{oscillating}, if the ratio $(f(n{+}1)/f(n))$ has no limit as $n\to\infty$. If, moreover, $\limsup_{n\to\infty}(f(n{+}1)/f(n))=\infty$ or $\liminf_{n\to\infty}(f(n{+}1)/f(n))=0$, then $f$ is said to be \textit{wild}. The oscillations of complexity for arbitrary, prefix-closed, and factorial regular languages (Problem~{\tt Reg2}) are described in

\begin{teo}[\cite{Sh08csr,Sh_r10ti}] \label{reg7}
All possible types of combinatorial complexity for arbitrary, prefix-closed, and factorial regular languages w.\,r.\,t. oscillation property are listed in the following table {\rm(}where W\,=\,wild, O\,=\,oscillating, N\,=\,non-oscil\-lating, $\alpha=\IR(L)$, $m=\PD(L)${\rm)}:\\[5pt]
\centerline{
\tabcolsep=4pt
\begin{tabular}{|l||l|l|l|l|}
\hline
Regular languages&$\alpha{=}1,m{=}0$&$\alpha{=}1,m{>}0$&$\alpha{>}1,m{=}0$&$\alpha{>}1,m{>}0$\\
\hline
\hline
Arbitrary&W,O,N&W,O,N&W,O,N&W,O,N\\
\hline
Prefix-closed&O,N&O,N&O,N&O,N\\
\hline
Factorial&O,N&N&O,N&O,N\\
\hline
\end{tabular}
}
\end{teo}

We finish the survey of Chapter 1 with the following property, which is distinctive for regular languages.

\begin{prp}[\cite{Sh08csr,Sh_r10ti}] \label{Tfact}
An arbitrary regular language $L$ has the same growth rate and polynomial index as its closures under taking prefixes, suffixes, and factors. Moreover, since such closures are not wild languages, each of them has the complexity $\Theta(n^{\PD(L)}\IR(L)^n)$.
\end{prp}

\subsection*{Chapter\,2 (\S\S\,5--11). Factorial languages. FAD-languages}

In this chapter we study general problems about factorial languages together with the problems about FAD-languages. First we describe the \textit{method of regular approximations} (\S\,5). Each factorial language $L$ over some alphabet $\Sigma$ has an antifactorial antidictionary $M=(\Sigma^*{-}L)\cup L\Sigma\cup\Sigma L$. We choose an arbitrary sequence $\{M_i\}$ of finite subsets of $M$ such that
$$
M_1\subseteq M_2\subseteq\ldots\subseteq M_i\subseteq \ldots\subseteq M,\ \ \bigcup_{i=1 }^{\infty}M_i=M
$$
(for instance, $M_i=M\cap\Sigma^{{\le}i}$). The FAD-languages $L_i$ with the antidictionaries $M_i$ are \textit{regular approximations} of $L$. We have
$$
L\subseteq\ldots\subseteq L_i\subseteq\ldots\subseteq L_1,\ \ \bigcap_{i=1 }^{\infty}L_i=L.
$$
One can check that $\lim_{i\to\infty}\IR(L_i)=\IR(L)$. By Theorem~\ref{reg3}, there is an algorithm that successively calculates, for any factorial language $L$, the members of a decreasing sequence that converges to $\IR(L)$. The convergence rate of such a sequence for some classes of factorial languages is really high, see below the results of Chapter 3.

To give a more detailed analisys of Problem {\tt Fact1}, we pay attention to the following questions. Let $L$ be an arbitrary factorial language. First, if $L$ has polynomial complexity, can the polynomial index of $L$ be found or approximated by means of regular approximations? Second, can one estimate the complexity of $L$ using some approximations of $L$ by regular languages from below? In an important particular case, the following proposition gives negative answers to both questions.

\begin{prp}[\cite{Sh06dlt}]\label{nopol}
If all words in an infinite factorial language $L$ are $\beta$-free for some number $\beta$, then all regular approximations of $L$ have exponential complexity and all regular subsets of $L$ are finite.
\end{prp}

Then we define and analyze \textit{FAD-automata}, which are ``canonical'' dfa's recognizing FAD-languages. FAD-automata are constructed from antidictionaries in linear time by a version of textbook Aho--Corasick's algorithm for pattern matching, see \cite{CMR}.

After this, we solve Problem {\tt Fact2} (\S\,6). As a target language, we take the \textit{Thue-Morse language} $\TM$. It consists of all factors of the Thue-Morse word, which is the fixed point of the binary morphism defined by the rule $\theta(a)=ab$, $\theta(b)=ba$. The following proposition describes the antidictionary of the Thue-Morse language. 

\begin{prp}[\cite{Sh_r05}]\label{tmad}
The antidictionary of the language $\TM$ is the set
\begin{gather*}
M=\{aaa,bbb\}\,\cup\,\{c\theta^i\!(aba)a,d\theta^i\!(bab)b,c\theta^i\!(bab)a,d\theta^i\!(aba)b\,|\ i\ge0\},\text{ where}\\
c,d\text{ are the last letters of } \theta^i(a) \text{ and }\theta^i(b) \text{, respectively}.
\end{gather*}
\end{prp}

The antidictionary $M$ contains words of length 3 and of length $3{\cdot}2^i{+}2$ for any $i\ge0$. Let $M_i=M\cap\{a,b\}^{\le3{\cdot}2^i{+}2}$ and additionally $M_{-1}=\{aaa,bbb\}$. The growth rates of the corresponding regular approximations are given by the following formula ($\phi$ denotes the golden ratio).

\begin{teo}[\cite{Sh_r05}]\label{tmappr}
Let $L_i$ be the FAD-language with the antidictionary $M_i$. Then $\IR(L_i)=\phi^{1/2^{i{+}1}}\!$.
\end{teo}

In \S\,7, we build two two-parameter series of FAD-automata: \textit{web-like} and \textit{generalized web-like} automata. They are used to prove Theorems~\ref{fadpol}--10. We call a language $L\in\Sigma^*$ \textit{symmetric} if it is closed under all automorphisms of the free monoid $\Sigma^*$. Problem {\tt FAD2} for the case of polynomial complexity is solved by

\begin{teo}[\cite{Sh09ita}] \label{fadpol} 
For any non-unary alphabet $\Sigma$ and any integer $m\ge0$, there exist both symmetric and non-symmetric FAD-languages over $\Sigma$ having the complexity $\Theta(n^m)$. 
\end{teo}

The following quite surprising theorem is proved in \S\,8. It shows that regular approximations of polynomial complexity cannot be used to find the polynomial index of the approximated language.

\begin{teo}[\cite{Sh07}] \label{appr2} 
For any non-unary alphabet $\Sigma$ and any integers $s$ and $m$ such that $1\le s\le m$, there exists a factorial language over $\Sigma$ having the complexity $\Theta(n^s)$ and such that almost all members of any sequence of its regular approximations have the complexity $\Theta(n^m)$. 
\end{teo}

\refstepcounter{teo}

Thus, the sequences of regular approximations of polynomial complexity have a ``non-compactness'' property: polynomial indices of approximations can stabilize arbitrarily far from the polynomial index of the target language. The only exception concerns the languages of bounded complexity.

\begin{prp}[\cite{Sh07}] \label{appr3} 
Almost all regular approximations of any factorial language of complexity $\Theta(1)$ have the complexity $\Theta(1)$.
\end{prp}

In \S\,9, we use the FAD-languages recognized by web-like and generalized web-like automata to approximate factorial languages from below. Thus, we answered the second question about the regular approximations in the affirmative, concluding the study of Problem {\tt Fact1}. Namely, we proved \textbf{Theorem \theteo} \cite{Sh09dam} stating that some languages defined by natural conditions have \textit{intermediate} (i.\,e., more than polynomial, but less than exponential) complexity. Factorial languages given by simple properties and having intermediate complexity are quite rare. So, the languages we have found are of certain interest. Let us describe one of two infinite series of such languages. 

The representation $w=a_1^{m_1}a_2^{m_2}\ldots a_t^{m_t}$, where $a_i\ne a_{i{+}1}$ for all $i$, is the \textit{power factorization} of a word $w$. The mentioned series consists of the languages of all words (over some fixed alphabet) satisfying the following two conditions on their power factorization:
\begin{itemize}
\item [--] the letters follow each other in accordance with some cyclic order;
\item [--] $m_1\le m_2\le\ldots\le m_{t{-}1}$.
\end{itemize}

In \S\,10, the transformations requested by Problem {\tt Fact3} are studied. Namely, we consider the restriction of the language to its \textit{extendable} (in one or both directions) part. The word $w\in L$ is \textit{two-sided extendable} in $L$ if there are arbitrarily long words $u$ and $v$ such that $uwv\in L$. The one-sided (say, right) extendability is defined in a similar way. The corresponding sets of extendable words are denoted by $\e(L)$ and $\re(L)$, respectively.

\begin{teo}[\cite{Sh06jm,Sh08ita}] \label{ext1} 
$\IR(L)=\IR(\re(L))=\IR(\e(L))$ for any factorial language $L$.
\end{teo}

Extendable parts of a language usually have simpler structure than the language itself. Hence, Theorem~\ref{ext1} can be useful for estimating the growth rates of factorial languages (e.\,g., we apply Theorem~\ref{ext1} to threshold languages in~\S\,16). On the other hand, more ``subtle'' parameters of complexity cannot be found from the extendable parts of a language, as the following theorem shows.
 
\begin{teo}[\cite{Sh06jm,Sh08ita}] \label{ext2} 
Each of the ratios $\C_L(n)/\C_{\re(L)}\!(n)$, $\C_L(n)/\C_{\e(L)}\!(n)$ can be a bounded, polynomial, or intermediate function.
\end{teo}

\refstepcounter{teo}

\S\,11, which is the last and biggest section of Chapter 2, is devoted to the FAD-languages of exponential complexity and their FAD-automata. The growth rate of a regular language $L$ (and, in many cases, other asymptotic parameters of complexity) can be found from a \textit{C-graph}, which is the subgraph generated by all nontrivial strong components of a consistent dfa recognizing $L$. So, C-graphs are the main objects of study in this section. Some of the results of this section are published in \cite{Sh_r09iu} while the others are contained in \cite{Sh11_1}.

We introduce two transformations of an antidictionary (\textit{reduction} and \textit{cleaning}). These transformations reduce the size of an antidictionary but preserve the $\Theta$-class of complexity of the factorial language with this antidictionary. This is exactly what is requested in Problem~{\tt FAD2}. Further, we build \textbf{Algorithm C} that decides whether a given arbitrary nontrivial strongly connected digraph is a strong component of some FAD-automaton and builds the corresponding FAD-language in the case of the affirmative answer. Thus, we get an algorithmic description of FAD-automata in terms of forbidden strong components (this result partially solves Problem {\tt FAD3}). Using this description, we enumerate all possible C-graphs (and hence, all possible growth rates) for the case of binary alphabet and ``small'' FAD-automata. This is our \textbf{Theorem \theteo} \cite{Sh_r09iu}, solving Problem {\tt FAD1} for a particular case. 

For bigger classes of FAD-languages the enumeration of growth rates is hardly possible, so a deciding algorithm would be the best solution we can hope for Problem {\tt FAD1}. On the base of Algorithm C we construct \textbf{Algorithm~G}, which builds the FAD-automaton having the same growth rate as the input strongly connected digraph and containing this digraph as a strong component. This algorithm allows one to construct a FAD-language with a given growth rate, thus providing a partial algorithmic solution to Problem~{\tt FAD1}. For a pity, we cannot use Algorithm G to prove that a given algebraic number is NOT a growth rate of a FAD-language. Namely, the following proposition suggests the idea that it is not possible to pick up a finite family of digraphs with the given index $\alpha$ such that the ``unsuccessful'' run of Algorithm G on all instances from this family proves that $\alpha$ is not a growth rate of a FAD-language.

\begin{prp} \label{fad3} 
An algebraic number $\alpha$ can simultaneously (a) be the growth rate of a FAD-language over an alphabet $\Sigma$ and (b) be the index of NO $r$-vertex digraph which is a strong component of a consistent dfa recognizing a FAD-language over $\Sigma$, where $r$ is the degree of $\alpha$. 
\end{prp}

In the end of \S\,11, we study the mutual location of strong components in FAD-automata. Recall that such a location determines the polynomial index of a regular language (Theorem~\ref{reg5}). The following propositions, proved by examples, show that from the complexity point of view FAD-languages form a quite representative subclass of $\tt Reg$.

\begin{prp} \label{fad4} 
$(1)$ There exist both symmetric and asymmetric binary FAD-languages having the complexity of type $\Theta(n\alpha^n)$ for some $\alpha>1$.\\
$(2)$ For any $k\ge3$, there exist $k$-ary FAD-languages having the complexity of type $\Theta(n^{k{-}2}\alpha^n)$ for some $\alpha>1$. 
\end{prp}

\begin{prp} \label{fad5} 
$(1)$ Over the binary alphabet, there exists a FAD-automa\-ton whose C-graph is not \emph{weakly} connected.\\
$(2)$ Over the binary alphabet, there exists a FAD-automaton whose nontrivial strong components form the $M_2$ poset w.\,r.\,t. reachability.
\end{prp}

\subsection*{Chapter 3 (\S\S\,12--18). Power-free languages}

This chapter is devoted to power-free languages except for \S\,18, in which we discuss the extension of our methods to pattern-free and Abelian power-free languages.

In \S\,12, we solve Problem {\tt PF1}. \textit{Exponent} of a word is the ratio between its length and its shortest period\footnote{The word of length $n$ over the alphabet $\Sigma$ can be seen as a function $w:\{1,\ldots,n\}\to\Sigma$. Periods of $w$ are the periods of this function.}. We call an exponent \textit{$k$-stable}, if there exists a $k$-ary word which has the exponent $\beta$ and is extendable to a double-infinite $\beta^+$-free word. The connection between $k$-stability and complexity is illustrated by the following note: if the exponent $\beta$ is not $k$-stable, then $\e(\L(k,\beta^+))=\e(\L(k,\beta))$ and hence $\IR(k,\beta^+)=\IR(k,\beta)$ by Theorem~\ref{ext1}. The following theorem shows that non-2-stable exponents clearly mark out the polynomial plateau (the exponents $\beta<2$ correspond to finite binary languages). Thus, Problem {\tt PF1} is solved.

\begin{teo}[\cite{Sh_r00}]\label{pfree1} 
The exponent $\beta$ is 2-stable if and only if $\beta=2$ or $\beta\ge7/3$. 
\end{teo}

\begin{cor} \label{pfree2} 
For any $\beta\in[2^+,7/3]$, the language $\L(2,\beta)$ has subexponential complexity.
\end{cor}

The last statement was first proved by Karhum\"aki and Shallit \cite{KS} (they even showed that all these languages have polynomial complexity). In addition, it was proved in \cite{KS} that the complexity of the language $\L(2,(7/3)^+)$ is exponential, that is, the polynomial plateau ends with the exponent $7/3$. But the latter result immediately follows from the fact that the language $\L(3,2)$ has exponential complexity \cite{Bra,Bri} and the following theorem.

\begin{teo}[\cite{Sh_r00}]\label{pfree3} 
There exists a morphism $f:\{1,2,3\}^*\to\{a,b\}^*$, mapping any square-free word to a $(7/3)^+$-free word.
\end{teo} 

The solution to Problem {\tt PF2} is given in \S\,13. A \textit{context} of a word $u$ in a language $L$ is a pair $(w_1,w_2)$ of words such that $w_1uw_2\in L$. By definition, the words $u,v\in L$ are \textit{context equivalent} if the sets of their contexts coincide. The corresponding decision problem  is called the \textit{context equivalence problem} (for the language $L$)\footnote{This problem is quite close to the word problem in the \textit{syntactic monoid} of the language $L$.}. This problem is little-studied and seems to be hard except for regular languages and the factorial languages satisfying the bounded gap property\footnote{A language $L$ satisfies the bounded gap property if there is a function $f(n)$ such that any word from $L$ of length $f(n)$ contains all words from $L$ of length $n$ as factors. See \cite{Kle} for the solution to the context equivalence problem for such languages.}. The solution to the context equivalence problem for the language $\L(2,2^+)$ of binary \textit{overlap-free} words is technically involved and reveals a non-trivial structure of this language. Nevertheless, the resulting \textbf{Algorithm E} is very fast.

\begin{teo}[\cite{Sh11sf}]\label{synt} 
The context equivalence of two arbitrary binary over\-lap-free words can be verified in the time linear in their total length.
\end{teo}
\refstepcounter{teo}
\refstepcounter{prp}

\begin{cor}[\cite{Sh11sf}]\label{csynt} 
The word problem in the syntactic monoid of the language $\L(2,2^+)$ can be solved in the time linear in the total length of input words. 
\end{cor}

The proof of Theorem~\ref{synt} is quite long. The main steps are 
\begin{itemize}
\item[--] a linear-time algorithm to check one-sided and two-sided extendability of an overlap-free word (\textbf{Theorem \theteo} \cite{Sh_r98});
\item[--] \refstepcounter{teo}\textbf{Proposition \theprp}\cite{Sh_r98}, stating non-equivalence of any nonequal two-sided extendable words;
\item[--] a necessary and sufficient condition of equivalence of one-sided extendable words (\textbf{Теорема \theteo}\cite{Sh_r98});
\item[--] the reduction of equivalence checking for nonextendable words to the comparison of finite sets of one-sided contexts \cite{Sh11sf};
\item[--] an algorithm that compares the sets of one-sided contexts of nonextendable words in linear time\cite{Sh11sf}.
\end{itemize}
The key role in the whole proof is played by the Thue--Morse morphism.

The studied problem has such a low time complexity mainly because of the extremely small set of binary overlap-free morphisms. Namely, the semigroup of all morphisms preserving the language $\L(2,2^+)$ is generated by the Thue-Morse morphism and the involution automorphism, see \cite{BS, See}. The same property holds for all binary power-free languages from the polynomial plateau \cite{Ram}. Hence, the above solution to the context equivalence problem can be applied, after a small correction, to any language from the polynomial plateau.

\medskip
\S\S\,14--15 are devoted to the development of the algorithms estimating complexity of power-free languages (Problem {\tt PF3}). An algorithm using regular approximations to obtain upper bounds for the growth rate of factorial languages should consist of three big steps:
\begin{itemize}
\item[1)] calculating the antidictionary of the chosen regular approximation;
\item[2)] building a consistent dfa from the antidictionary;
\item[3)] calculating the growth rate of the regular approximation from the dfa.
\end{itemize}
An efficient impementation of steps 2 and 3 is provided by the mentioned above version of Aho--Corasick's algorithm and by Algorithm R, respectively. If we perform step 1 by some optimized exhaustive search, we will obtain an algorithm \cite{Sh08csr} which allows one to get much better upper bounds than the algorithms described in  \cite{NZ,OR}. The main flaw of this straightforward algorithm follows from the type of dependence of the time and space expences on the alphabet: both these expences include the factor $k!$, where $k$ is the alphabetic size. So, it is hardly possible to proceed languages over more than 4--5 letters.

The main result of \S\,14 is \textbf{Algorithm U} \cite{Sh10tcs} for upper bounds. The time and space required by Algorithm U to process a language are approximately $k!$ times less than the time and space used by the straightforward algorithm. Such a gain is obtained by using symmetry of power-free languages. Instead of a FAD-automaton, the algorithm directly builds a ``factor automaton'' which has the same index but much less size. The practical efficiency of Algorithm U is demonstrated, for example, by Table~\ref{tablow}, while the theoretical one is described by the following

\begin{teo}[\cite{Sh10tcs}] \label{upper1} 
Suppose that $M$ is the antidictionary of the language $\L(k,\beta)$, $M_m$ is its subset consisting of all words of period at most $m$, and $N=(m\beta \C_{\L(k,\beta)}(m))/k!$. Then the factor automaton whose index is equal to the growth rate of the regular approximation $\L_m(k,\beta)$ has $O(N)$ vertices and can be constructed from the triple $(m,\beta,k)$ in time $O(N\log N)$ and space $O(N)$. 
\end{teo}

To build the antidictionary in an optimized way, we use 

\begin{teo}[\cite{Sh11_2}] \label{upper2} 
Let $1<\beta<2$, $xy\in \L(k,\beta)$. If the $\beta$-power $(xy)^\beta$ is not minimal, then the word $(xy)^\beta$ contains a $\beta$-power $(zt)^{\beta}$ such that $|(zt)^{\beta}|<|xy|$. Moreover, if $\beta\le(4/3)^+$, then $|zt|\le|y|$, and if $\beta\le(5/4)^+$, then $|zt|<|y|$.
\end{teo}

Lower bounds for the growth rates of power-free languages cannot be obtained similarly to the upper bounds, see Proposition~\ref{nopol}. But in the case $\beta\ge2$, it is possible to use the properties of factor automata to convert the upper bounds obtained from them to the two-sided bounds. 

\begin{teo}[\cite{Sh09dlt}]\label{lower1} 
Suppose that $\beta\ge2$, $k$ and $m$ are positive integers, $M_m$ is the set of all words of period $\le m$ from the antidictionary of the language $\L(k,\beta)$, $L_m$ is the regular approximation of $\L(k,\beta)$ with the antidictionary $M_m$, and the FAD-automaton recognizing $L_m$ has a unique nonsingleton strong component\footnote{Algorithm R performs splitting of the processed dfa into strong components. So, the latter condition is already checked during the calculation of the growth rate of $L_m$. It seems probable that this condition is always satisfied, but it is not proved yet.}. Then any number $\gamma$ such that $\gamma+\frac{1}{\gamma^{m{-}1}(\gamma{-}1)}\le\IR(L_m)$ satisfies the inequality $\gamma<\IR(\L(k,\beta))$.
\end{teo}

The idea of such a conversion of upper bounds into the two-sided ones was suggested by Kolpakov. In \cite{Kol,Kol07} he obtained good enough lower bounds for the growth rates of $\L(3,2)$ and $\L(2,3)$. But the method of Kolpakov is not universal (one should derivate approximating formulas for each language separately) and uses quite time-consuming procedures. Our method is free from these flaws. It is absolutely universal, because $k$ and $\beta$ are not used in the calculation of $\gamma$; only the numbers $m$ and $\IR(L_m)$, provided by Algorithm U, are needed. In addition, $\gamma$ can be calculated with any precision in an almost constant time. In the particular cases considered by Kolpakov our bounds are much more precise.

The computer implementation of Algorithm U (together with the attachment for calculating lower bounds) allowed us to considerably improve all known bounds of the growth rates for power-free languages and to obtain lots of previously unknown bounds. Selected results are given in Table~\ref{tablow}\footnote{Tables with numerical bounds for the growth rates of different power-free languages can be also found in \cite{Sh10tcs,Sh09dlt,ShGo10}.}. All of them are obtained using a PC with a 3.0GHz CPU and 2Gb of memory. All bounds are rounded off to 7 digits after the dot. If only one bound is given, then these digits are the same for both lower and upper bounds.

\begin{table}[!htb]
\caption{\small\sl Bounds for the growth rates of $\beta$-free languages with $\beta\ge2$.} \label{tablow}
\vspace*{2mm}
\tabcolsep=2.7pt
{\footnotesize\centerline{ 
\begin{tabular}{|l|l|c|}
\hline
$k$&$\beta$&оценки\\
\hline
2&$\!(7/3)^+\!$&1.2206318--1.2206448\\
2&$\!(5/2)^+\!$&1.3662971--1.3663011\\
2&$3$&1.4575732--1.4575773\\
2&$3^+$&1.7951246--1.7951264\\
2&$4$&1.8211000\\
2&$4^+$&1.9208015\\
3&$2$&1.3017597--1.3017619\\
3&$2^+$&2.6058789--2.6058791\\
3&$3$&2.7015614--2.7015616\\
3&$3^+$&2.9119240--2.9119242\\
3&$4$&2.9172846\\
3&$4^+$&2.9737546\\
\hline
\end{tabular} 
\hspace*{1.6mm}
\begin{tabular}{|r|r|r|r|r|}
\hline
$k\big\backslash\,\beta$&$2$&$2^+$&$3$&$3^+$\\
\hline
4&2.6215080&3.7284944&3.7789513&3.9487867\\
5&3.7325386&4.7898507&4.8220672&4.9662411\\
6&4.7914069&5.8277328&5.8503616&5.9760100\\
7&5.8284661&6.8537250&6.8705878&6.9820558\\
8&6.8541173&7.8727609&7.8858522&7.9860649\\
9&7.8729902&8.8873424&8.8978188&8.9888625\\
10&8.8874856&9.8988872&9.9074705&9.9908932\\
11&9.8989813&10.9082635&10.9154294&10.9924142\\
12&10.9083279&11.9160348&11.9221106&11.9935831\\
13&11.9160804&12.9225835&12.9278022&12.9945010\\
14&12.9226167&13.9281788&13.9327109&13.9952350\\
15&13.9282035&14.9330157&14.9369892&14.9958311\\
\hline
\end{tabular} }
\vspace*{2mm}
\hspace*{1.1cm}Small alphabets\hspace*{4.7cm}Large alphabets}
\vspace*{-0.9mm}
\end{table}

Among all power-free languages, threshold languages are the most interesting from the complexity point of view. From the obtained numerical results, it clearly follows that the sequences of growth rates of regular approximations for threshold languages demonstrate the slowest convergence among all such sequences for power-free languages. Threshold languages are considered in \S\,16. To study them, we introduce the notion of \textit{$m$-repetition}, which is any word $u^{\beta}$ belonging to the antidictionary of the considered language and satisfying the condition $|u^{\beta}|-|u|=m$. For a threshold language $\L(k,\beta)$, we denote by $\L^{(m)}(k)$ its regular approximation, whose antidictionary consists of all $r$-repetitions with $r\le m$. It is not hard to see that the growth rates of all languages $\L^{(2)}(k)$ coincide. Close similarity of the structure of these languages can be observed using \textit{cylindric representation} of words \cite{ShGo08,ShGo10}. Such a similarity also takes place for more precise regular approximations, as the following theorem shows.

\begin{teo}[\cite{ShGo10}]\label{dej}
For any fixed integer $m\ge3$, there exists a finite set $D_m$ of ternary two-dimensional words of size $O(m)\times O(m)$ such that for any $k>2m{-}3$ a word belongs to the language $\L^{(m)}(k)$ if and only if its cylindric representation has no factors from the set $D_m$.
\end{teo}

We calculate the sets $D_m$ for $m=3,4,5,6$ (\!\cite{ShGo08,ShGo10}; the set $D_7$, calculated by Gorbunova, also can be found in \cite{ShGo10}). Using these sets, we calculate the growth rates of the languages $\L^{(m)}(k)$ for different $k$. Analyzing both these results and the results obtained by the direct use of Algorithm U we formulate the following conjecture.

\begin{conj}[\cite{ShGo08}; revised in \cite{ShGo10}] \label{dej3}
The sequence of growth rates of $k$-ary threshold languages converges to a limit $\hat\alpha\approx1.242$ as $k$ approaches infinity.
\end{conj}

Conjecture \ref{dej3} naturally follows from the above description of threshold languages (which is our solution to Problem {\tt PF5}). It strengthens Dejean's conjecture\footnote{Conjecture \ref{dej3} was published before Dejean's conjecture was proved.} and refutes the idea that the growth rates of threshold languages tend to 1 as the alphabets increase. Currie and Rampersad \cite{CR2} mention that when proving Dejean's conjecture they obtained some results supporting Conjecture~\ref{dej3}.

\smallskip
Using Algorithm U and Theorem \ref{lower1} we obtain numerical bounds for the growth rates of the languages $\L(k,\beta)$ for a wide range of alphabets and exponents. As a result, we are able to represent the behaviour of the growth rate as a function $\alpha(k,\beta)$. The empirical laws of behaviour of this function are presented and then explained in \S\,17. We derive several asymptotic formulas for $\alpha(k,\beta)$, thus solving Problem {\tt PF4}. For the case $\beta>2$ one has

\begin{teo}[\cite{Sh10csr}] \label{big}
Let $\beta\in[n^+,n{+}1]$, where $n\ge2$ is an integer. Then
\begin{equation*}
\alpha(k,\beta)=
\left\{
\begin{array}{ll}
k-\frac{1}{k^{n{-}1}}+\frac{1}{k^n}-\frac{1}{k^{2n{-}2}}+O\big(\frac{1}{k^{2n{-}1}}\big),&\text{\rm if\ } \beta\in[n^+,n{+}\frac{1}{2}],\\
k-\frac{1}{k^{n{-}1}}+\frac{1}{k^n}+O\big(\frac{1}{k^{2n{-}1}}\big),&\text{\rm if\ } \beta\in[(n{+}\frac{1}{2})^+,n{+}1].
\end{array}
\right. \label{formbig}
\end{equation*}
\end{teo}

\begin{cor}[\cite{Sh10csr}]\label{appr_k}
For any fixed $\beta\ge2^+$, the difference $(k-\alpha(k,\beta))$ approaches zero at polynomial rate as $k\to\infty$. For any fixed $k\ge2$, the same difference approaches zero at exponential rate as $\beta\to\infty$.
\end{cor}

\begin{cor}[\cite{Sh10csr}]\label{jumps_big}
For a fixed $k$, the jumps of the function $\alpha(k,\beta)$ at the endpoints of the interval $[n^+,n{+}1]$ are much bigger than the variation of this function inside this interval. Namely,
\begin{align*}
\alpha(k,n^+)-\alpha(k,n)&=\tfrac{1}{k^{n{-}2}}+O\big(\tfrac{1}{k^{n{-}1}}\big),\\ \alpha(k,n{+}1)-\alpha(k,n^+)&=\tfrac{1}{k^{2n{-}2}}+O\big(\tfrac{1}{k^{2n{-}1}}\big). 
\end{align*}
\end{cor}

Next we analyze the behaviour of $\alpha(k,\beta)$ at the point $\beta=2$.

\begin{prp}[\cite{Sh10csr}]\label{two}
The following equalities hold:
\begin{align*}
\alpha(k{+}1,2)&=k-\tfrac{1}{k}-\tfrac{1}{k^3}+O\big(\tfrac{1}{k^5}\big);\\
\alpha(k,2^+)&=k-\tfrac{1}{k}-\tfrac{1}{k^3}-\tfrac{1}{k^4}+O\big(\tfrac{1}{k^5}\big).
\end{align*}
\end{prp}

\begin{cor}[\cite{Sh10csr}]\label{jump}
For any $k$, the function $\alpha(k,\beta)$ jumps by more than a unit at the point $\beta=2$. Namely, $\alpha(k,2^+)-\alpha(k,2)=1+\frac{1}{k^2}+O(\frac{1}{k^3})$.
\end{cor}

\begin{cor}[\cite{Sh10csr}]\label{backw}
At any point $(k,2)$, the increment of $k$ by 1 and the addition of $\,^+\!$ to the exponent almost equally affect the growth rate of the power-free language. Namely, $\alpha(k{+}1,2)-\alpha(k,2^+)=\tfrac{1}{k^4}+O\big(\tfrac{1}{k^5}\big).$
\end{cor}

All asymptotic formulas given above work perfectly even for small alphabets, predicting the values from Table~\ref{tablow} with a good precision. For $\beta<2$, our main results are the following two conjectures. They are based on a number of partial results and numerical bounds.

\begin{conj}[\cite{Sh10csr}]\label{avec}
The following equalities hold for any fixed integers $n,k$ such that $k>n\ge3$: 
\begin{equation*}
\arraycolsep=2pt
\begin{array}{lcl}
\alpha(k,\frac{n}{n{-}1}\!^+)&=&k{+}2{-}n{-}\frac{n{-}1}{k}{+}O\big(\frac{1}{k^2}\big),\\ \alpha(k,\frac{n}{n{-}1})&=&k{+}1{-}n{-}\frac{n{-}1}{k}{+}O\big(\frac{1}{k^2}\big).
\end{array}
\end{equation*}
\end{conj}

Conjecture~\ref{avec} predicts that the properties found above for the point $\beta=2$ hold true for all points $\beta=\frac{n}{n{-}1}$ such that $2<n<k$. Indeed, Conjecture~\ref{avec} implies

\begin{cor}[\cite{Sh10csr}]
Let $n$ and $k$ be integers such that $2<n<k$. Then
\begin{equation*}
\arraycolsep=2pt
\begin{array}{lclclcl}
\alpha(k,\frac{n}{n{-}1}^+)&-&\alpha(k,\frac{n}{n{-}1})&=&1&+&O\big(\frac{1}{k^2}\big)\\
\alpha(k,\frac{n}{n{-}1})&-&\alpha(k,\frac{n{+}1}{n}^+)&=&\frac{1}{k}&+&O\big(\frac{1}{k^2}\big)\\
\alpha(k{+}1,\frac{n}{n{-}1})&-&\alpha(k,\frac{n}{n{-}1}^+)&=&&&O\big(\frac{1}{k^2}\big).
\end{array}
\end{equation*}
\end{cor}

The second conjecture describes the behaviour of the function $\alpha(k,\beta)$ for the case when $\beta$ depends on $k$ such that the obtained language is close to a threshold language.

\begin{conj}[\cite{Sh10csr}]\label{smal}
For any integer $n\ge0$ the limits 
$$
\alpha_n=\lim\limits_{k\to\infty}\alpha(k,{\tfrac{k{-}n}{k{-}n{-}1}\!}^+)\text{\rm\ \ and\ \ } \alpha_n'=\lim\limits_{k\to\infty}\alpha(k,\tfrac{k{-}n}{k{-}n{-}1})
$$
exist. Moreover, $\alpha_{n{+}1}'=\alpha_n\ $ and $\ \alpha_{n{+}1}-\alpha_n>1$.
\end{conj}

Note that $\alpha_0\approx1.242$ according to Conjecture \ref{dej3}.

\medskip
In \S\,18, we demonstrate how our methods for power-free languages can be extended to estimate the growth rates of pattern-free and Abelian power-free languages. Recall that if $u$ is a word, then a word $w$ is said to \textit{avoid the pattern} $u$ if there are no homomorphic images of $u$ among the factors of $w$. An $P$-free language consists of all words (over a given alphabet) avoiding all patterns from the set $P$. Abelian powers generalize ordinary powers: two words are considered equal if they are anagrams of each other. Abelian power-free languages are defined in the same way as power-free languages.

All pattern-free and Abelian power-free languages are symmetric. As a result, we can apply Algorithm U to such languages in order to get upper bounds for their growth rates. Only the procedure building the antidictionary should be appropriately changed. We develop such a universal procedure for Abelian power-free languages in a joint work with Samsonov \cite{SaSh}. On the other hand, such procedures for pattern-free languages heavily depend on the avoided patterns. So, we focused on two particular binary languages; they avoid two very similar sets $P_1=\{xxyxxy,xxx\}$ and $P_2=\{xyxxyx,xxx\}$ respectively. For the language avoiding $P_1$ we adopt Theorem~\ref{lower1} to find a sharp two-sided bound of the growth rate. This rate is about $1.0989$. In contrast, we prove, using a modification of Algorithm U, that the second mentioned language is finite (and then has zero growth rate). These results are contained in \cite{Sh11_3}.

\subsection*{Chapter 4 (\S\S\,19--20). Languages of minimal powers}

This chapter is closely connected to the previous one, because the study of factorial languages is impossible without paying attention to their antidictionaries. A word $w$ from the antidictionary of a factorial language $L$ has nearly the same structure as the words from $L$, because all proper factors of $w$ belongs to $L$. On the other hand, the antidictionary of $L$ has the structure completely different from that of $L$. Obviously, both similarity and difference affect complexity. 

Complexity of any language of minimal powers has some ``trivial'' zeroes (for example, a square cannot have odd length). In order to exclude trivial zeroes from consideration, we introduce a version of combinatorial complexity called \textit{root complexity}. The root complexity $R_{k,\beta}(n)$ of the language of minimal $k$-ary $\beta$-powers returns the number of such powers of \textit{period} $n$. If $u^\beta$ is a minimal $\beta$-power, then the word $u$ is $\beta$-free. Hence, the complexity of a power-free language exceeds the root complexity of its antidictionary. But the numerical results show that the growth rates of these two complexities are very close for any power-free language.

The root complexity of an antidictionary behaves much less regular than the complexity of the corresponding factorial language. That is why we study root complexity mostly within the bounds of Problem {\tt MP}. Zeroes of the function $R_{k,\beta}(n)$ are exactly the ``forbidden'' periods for minimal $k$-ary $\beta$-powers.

Problem 1.12 of \cite{AS} asks about zeroes and behaviour of the function $R_{3,2}(n)$. We study this problem in \S\,19. Minimal squares are closely connected to square-free \textit{circular words} as the following proposition shows. Recall that a circular word is just a cyclic sequence of letters; the factors of a circular word are usual words. Any word can be transformed to its \textit{circular closure} by linking up the ends together.

\begin{prp}[\cite{Sh10dlt,Sh10ejc}] \label{psqf1}
A word of the form $u^2$ is a minimal square if and only if the circular closure of $u$ is square-free. 
\end{prp}

Due to Proposition~\ref{psqf1}, we formulate the main result of \S\,19 both in terms of circular words and in terms of root complexity.
 
\begin{teo}[\cite{Sh10ejc}] \label{sf}
$(1)$ A ternary square-free circular word of length $n$\\
\indent$(1\mathrm a)$ exists if and only if $n\notin\{5,7,9,10,14,17\};$\\
\indent$(1\mathrm b)$ is unique\footnote{up to renaming the letters.} if and only if $n\in\{1,2,3,4,6,8,11,12,13,15,16,21\}$.\\
$(2)$ The number of ternary square-free circular words of length $n$ depends on $n$ exponentially.
\end{teo}

\begin{cor} \label{csqf1}
The function $R_{3,2}(n)$ is exponential. Moreover,
$$
R_{3,2}(n)=
\begin{cases}
0,& \text{if\ }\ n=5,7,9,10,14,17,\\
3,& \text{if\ }\ n=1,\\
6,& \text{if\ }\ n=2,3,4,6,8,11,12,13,15,16,21,
\end{cases}
$$
and $R_{3,2}(n)\ge12$ otherwise.
\end{cor}

The statement (1a) of Theorem \ref{sf} was first proved by Currie \cite{Cur} with the aid of relatively long computer search. As a result, the proof by Currie cannot clarify the structure of ternary square-free circular words or the nature of the exceptions found. We give a computer-free proof, revealing an interesting connection between ternary square-free circular words and closed walks in the weighted $K_{3{,}3}$ graph. All statements of Theorem \ref{sf} are proved in parallel and all exceptions are made visual.

\medskip
The dissertation is concluded by \S\,20, in which zeroes of an arbitrary function $R_{k,\beta}(n)$ are described. The results are formulated in terms of permitted/forbidden periods of minimal powers. Recall that words $u$ and $v$ are \textit{conjugates} if $u=yz$ and $v=zy$ for some $y$ and $z$. Minimal powers over the binary alphabet are described in the following theorem.

\begin{teo}[\cite{Sh10dlt}] \label{bin1}
A binary minimal $\beta$-power of period $p$\\
$(1)$ exists for any positive integer $p$ if $\beta\ge(5/2)^+;$\\
$(2)$ exists for any positive integer $p\notin\{5,9,11,17,18\}$ if \hbox{$\beta\in[(7/3)^+\!,5/2];$}\\
$(3)$ is a power of a conjugate of the word $\theta^m(a)$, $\theta^m(b)$, $\theta^m(aba)$, or $\theta^m(bab)$ for some $m\ge0$, where $\theta$ is the Thue--Morse morphism, if $\beta\in[2^+\!,(7/3)];$ in particular, $p=2^m$ or $p=3\cdot2^m$.
\end{teo}

Proving the second statement of Theorem~\ref{bin1}, we finalize the description of possible lengths of binary $\beta$-free circular words (all other values of $\beta$ were studied by Aberkane and Currie \cite{AC1,AC2}). Note that the lists of exceptions in statement (2) of Theorem~\ref{bin1} and in the following corollary are slightly different.

\begin{cor}[\cite{Sh10dlt}]\label{ccycl}
Let $\beta\in[(7/3)^+\!,5/2]$. The binary $\beta$-free circular word of length $n$ exists if and only if $n\notin\{5,9,11,18\}$.
\end{cor}

Next we move to the alphabets with more than two letters.

\begin{teo}[\cite{Sh10dlt}] \label{big4}
Any positive integer is a period of some minimal \mbox{$k$-ary} $\beta$-power if one of the following conditions holds:\\
$(1)$ $k\ge4$ and $\beta=2;$  $(2)$ $k\ge3$ and $\beta\ge2^+;$  $(3)$ $\beta>\RT(\lfloor k/2\rfloor)$.
\end{teo}

On the other hand, forbidden periods exist when $\beta\approx\RT(k)$.  Some of them are listed in the following theorem.

\begin{teo}[\cite{Sh10dlt}] \label{small1}
There exist no minimal $k$-ary $\beta$-power of period $p$ if one of the following conditions hold:\\
$(1)\ \beta\in\big[{\frac{k}{k{-}1}\!}^+,\frac{k{-}1}{k{-}2}\big]$ and $p$ satisfies one of the restrictions\\
\indent $\mathrm (a)$ $k<p<\big\lceil\frac{k{+}1}{2}\big\rceil(k{-}1)\,$ and $\,p\bmod k\ne0$,\\
\indent $\mathrm (b)$ $p\in[(m{-}2)(k{+}1){+}1,m(k{-}1){-}1]$ for some integer $m\ge\big\lceil\frac{k{+}3}{2}\big\rceil$ and $\,p\bmod k\ne0$,\\
\indent $\mathrm (c)$ $p=3k$ or $p=4k;$\\
$(2)\ \beta\in\big[\frac{k{-}1}{k{-}2}^+\!,\frac{k{-}2}{k{-}3}\big]$ and $p\in[(m{-}1)(k{+}1){+}1,m(k{-}2){-}1]$ for some integer $m\in[2,k{-}2];$\\   
$(3)\ k\ge9$, $\beta\in\big[\frac{2k{-}5}{2k{-}7}^+\!,\frac{k{-}3}{k{-}4}\big]$, and $p=2k{-}7$. 
\end{teo}

\begin{cor}[\cite{Sh10dlt}]
For $\beta\in\big[{\frac{k}{k{-}1}\!}^+,\frac{k{-}1}{k{-}2}\big]$, and also for $\beta\in\big[\frac{k{-}1}{k{-}2}^+\!,\frac{k{-}2}{k{-}3}\big]$ in the case $k\ge7$, the minimal $k$-ary $\beta$-powers of period $p$ do not exist for $\Omega(k^2)$ different values of $p$. 
\end{cor}

Since the existence of a minimal $k$-ary $\beta$-power of period $p$ is decidable for any fixed triple $(k,\beta,p)$, we can add the results of computer check to the above theorems. Finally, we get the following general conjecture about the existence and distribution of forbidden periods.

\begin{conj}[\cite{Sh10dlt}] \label{forper}
Let $k\ge3$, $\beta>\RT(k)$.\\
$(1)$ For a pair $(k,\beta)$, there exists a forbidden period if and only if one of the following conditions is satisfied:\\
\indent $\mathrm (a)$ $\beta\le\frac{k{-}1}{k{-}2};$\\
\indent $\mathrm (b)$ $k=6$ and $\beta\in[\frac{9}{7}^+\!,\frac{4}{3}]$, or $k\ge7$ and $\beta\in\big[\frac{k{-}1}{k{-}2}^+\!,\frac{k{-}2}{k{-}3}\big];$\\
\indent $\mathrm (c)$ $k\ge9$ and $\beta\in\big[\frac{2k{-}5}{2k{-}7}^+\!,\frac{k{-}3}{k{-}4}\big]$.\\
$(2)$ For any pair $(k,\beta)$, the set of forbidden periods is finite.\\
$(3)$ If $k\ge9$, then any period $p\ge k(k{-}1)$ is permitted for any pair $(k,\beta)$.
\end{conj}

We conclude with two short comments on this conjecture. First, the intervals for $\beta$ mentioned in statement (1) coincide with such intervals mentioned in Theorem~\ref{small1}. Second, the bound on $p$ in statement (3) is the best possible, because the period $k(k{-}1){-}1$ is forbidden for any $\beta\le\frac{k{-}1}{k{-}2}$ by Theorem~\ref{small1}\,(1).

\end{document}